# Spin Torque Ferromagnetic Resonance Induced by the Spin Hall Effect

Luqiao Liu, Takahiro Moriyama, D. C. Ralph, and R. A. Buhrman

Cornell University, Ithaca, New York, 14853

## ABSTRACT

We demonstrate that the spin Hall effect in a thin film with strong spin-orbit scattering can excite magnetic precession in an adjacent ferromagnetic film. The flow of alternating current through a Pt/NiFe bilayer generates an oscillating transverse spin current in the Pt, and the resultant transfer of spin angular momentum to the NiFe induces ferromagnetic resonance (FMR) dynamics. The Oersted field from the current also generates an FMR signal but with a different symmetry. The ratio of these two signals allows a quantitative determination of the spin current and the spin Hall angle.

The spin Hall effect (SHE), the conversion of a longitudinal charge current density $J_C$ into a transverse spin current density $J_S \hbar/2e$, originates from spin-orbit scattering [1-4], whereby conduction electrons with opposite spin orientations in a nonmagnetic metal [5] or semiconductor [6] are deflected in opposite directions. The SHE has attracted widespread interest because it can generate pure spin currents from a nonmagnetic source, a phenomenon that could find important applications in future spintronic devices. Several techniques [5, 7, 8] have been developed to determine the magnitude of the SHE, which is generally characterized by the spin Hall angle, $\theta_{SH} = J_S/J_C$. For thin-film Pt, estimates of $\theta_{SH}$ obtained using different approaches differ by more than an order of magnitude [8-10], but already there have been efforts to utilize the spin current that arises from the SHE, first to tune the damping coefficient in a ferromagnetic metal [8], and, most recently, to induce a spin wave oscillation in a ferrimagnetic insulator having small damping [11]. Here we show that the SHE can also be used to excite dynamics in an ordinary metallic ferromagnet. Our experiment also allows a quantitative determination of the SHE strength that is self-calibrated, as explained below, enabling measurements of the spin currents generated by the SHE with small experimental uncertainties.

We study Pt/Permalloy bilayer films with a microwave-frequency (RF) charge current applied in the film plane (Permalloy = Py = $Ni_{81}Fe_{19}$). An oscillating transverse spin current is generated in the Pt by the SHE and injected into the adjacent Py (Fig. 1(a)), thereby exerting an oscillating spin torque (ST) on the Py that induces magnetization precession. When the frequency and field bias satisfy the FMR condition for the Py, strong resonant precession results in a significant oscillation of the bilayer resistance due to the anisotropic magnetoresistance (AMR) of the Py. This generates a DC voltage signal across the sample from the mixing of the RF current and the oscillating resistance, similar to the signal that arises from ST induced FMR in spin valves and magnetic tunnel junctions [12-15]. The resonance properties enable a direct quantitative measure of the spin current absorbed by the Py.

Our measurement setup is shown in Fig. 1(c). Pt/Py bilayers were grown by DC magnetron sputter deposition. The individual layer thicknesses were 4-15 nm, with specific values stated below. The starting material for the Pt was 99.95% pure. Highly resistive Ta (1 nm) was employed as the capping layer to prevent oxidation of the Py. The bilayers were subsequently patterned into microstrips using photolithography and ion milling. The samples' widths ranged from 1 to 20 μm and the lengths from 3 to 250 μm. By using a bias tee, we were able to apply a microwave current to our sample and at the same time measure the DC voltage. A sweeping magnetic field $H_{ext}$ was applied in the film plane, with the angle $\theta$ between $H_{ext}$ and microstrip kept at 45° unless otherwise indicated. The output power of the microwave signal generator was varied from 0 to 20 dBm and the measured DC voltage was proportional to the applied power, indicating that the induced precession was in the small angle regime. All the measurements we present were performed at room temperature with a power of 10 dBm.

We model the motion of the Py magnetic moment $\hat{m}$ by the Landau-Lifshitz-Gilbert equation containing the ST term [16]:

$$\frac{d\hat{m}}{dt} = -\gamma \hat{m} \times \vec{H}_{eff} + \alpha \hat{m} \times \frac{d\hat{m}}{dt} + \gamma \frac{\hbar}{2e\mu_0 M_s t} J_{S,RF} (\hat{m} \times \hat{\sigma} \times \hat{m}) - \gamma \hat{m} \times \vec{H}_{RF}. \qquad (1)$$

Here $\gamma$ is the gyromagnetic ratio, $\alpha$ is the Gilbert damping coefficient, $\mu_0$ is the permeability in vacuum, $M_s$ is the saturation magnetization of Py, $t$ is the thickness of the Py layer, $J_{S,RF} \hbar/2e$ represents the oscillating spin current density injected into Py, $H_{RF}$ is the Oersted field generated by the RF current, $H_{eff}$ is the sum of $H_{ext}$ and the out-of-plane demagnetization field $4\pi M_{eff}$, and $\hat{\sigma}$ is the direction of the injected spin moment. The third and fourth terms on the right hand side of Eq. (1) are the result of in-plane spin torque and the out-of-plane torque due to the Oersted field, respectively (Fig. 1(a)). The magnetic-resonance mixing signal in response to a combination of in-plane and out-of-plane torques has been calculated in the context of ST-driven FMR [14, 15], which we can translate to our notation as:

$$V_{mix} = -\frac{1}{4}\frac{dR}{d\theta}\frac{\gamma I_{RF}\cos\theta}{\Delta 2\pi (df/dH)|_{H_{ext}=H_0}}[SF_S(H_{ext}) + AF_A(H_{ext})], \qquad (2)$$

where $F_S(H_{ext}) = \Delta^2/[\Delta^2 + (H_{ext}-H_0)^2]$ is a symmetric Lorentzian function centered at the field $H_0$ with linewidth $\Delta$, $F_A(H_{ext}) = F_S(H_{ext})(H_{ext}-H_0)/\Delta$ is an antisymmetric Lorentzian, $S = \hbar J_{S,RF}/(2e\mu_0 M_s t)$, $A = H_{RF}[1+(4\pi M_{eff}/H_{ext})]^{1/2}$, $R$ is the resistance of the stripline, $I_{RF}$ is the microwave current through the stripline, and $f$ is the resonance frequency. We therefore expect the resonance signal to consist of two parts, a symmetric Lorentzian peak proportional to the spin current density and an antisymmetric peak proportional to $H_{RF}$.

The Oersted field $H_{RF}$ can be calculated from the geometry of the sample. Since the microwave skin depth is much greater than the Py thickness the current density in the Py should be spatially uniform, and in this case the Oersted field from the charge current density in the Py should produce no net torque on the Py [see Fig. 1(b)]. The Oersted field can therefore be calculated entirely from the current density $J_{C,RF}$ in the Pt layer. The microstrip width is much larger than the Pt thickness, so the sample can be approximated as an infinitely wide conducting plate and the Oersted field determined by Ampère's law, $H_{RF} = J_{C,RF}d/2$, where $d$ is the Pt thickness. We checked $H_{RF}$ by numerical integration and the difference is less than 0.1% from the infinite plate approximation. Using this result, the ratio of the spin current density entering the Py to the charge current density in the Pt can then be determined quantitatively in a simple way from the ratio of the symmetric and antisymmetric components of the resonance curve

$$\frac{J_{S,RF}}{J_{C,RF}} = \frac{S}{A}\frac{e\mu_0 M_s td}{\hbar}\left[1+\left(4\pi M_{eff}/H_{ext}\right)\right]^{1/2}. \quad (3)$$

All of the parameters entering Eq. (3) are either fundamental constants or quantities that can be measured directly, so this expression allows a measurement of $J_{S,RF}/J_{C,RF}$ with small experimental uncertainties. The measurement is self-calibrated in the sense that the strength of the torque from the spin current is measured relative to the torque from $H_{RF}$, which can be calculated easily from the geometry of the sample.

Figure 2(a) shows the ST-FMR signals measured on a Pt(6)/Py(4) (thicknesses in nanometers) sample for $f$ = 5-10 GHz. As expected from Eq. (2), the resonance peak shapes can be very well fit by the sum of symmetric and antisymmetric Lorentzian curves with the same linewidth for a given $f$ (fits are shown as lines in Fig. 2(a)). The fact that the symmetric peak changes its sign when $H_{ext}$ is reversed (inset of Fig. 2(a)) agrees with the form of spin torque $\vec{\tau}_{ST} \propto \hat{m}\times\hat{\sigma}\times\hat{m}$ given in Eq. (1), and excludes the possibility that the signal is due to an unbalanced perpendicular Oersted field torque, in direction $\hat{m}\times\hat{H}_{RF}^{\perp}$, which would yield symmetric peaks with the same sign for opposite $H_{ext}$. The resonant peak positions are summarized in Fig. 2(b), and agree well with the Kittel formula

$f = (\gamma/2\pi)\left[H_0\left(H_0+4\pi M_{eff}\right)\right]^{1/2}$. From a one-parameter fit to the resonance frequencies we

An additional contribution to the DC voltage can arise from spin pumping by the precessing moment in combination with the inverse SHE in the Pt layer, as observed in Ref. 10. However, this effect is second order in $\theta_{SH}$ in our geometry and we calculate that it should contribute a negligible voltage, about two orders of magnitude smaller than the signals we measure.

determine that the demagnetization field $4\pi M_{eff} = 0.805 \pm 0.005$ T for the Pt(6)/Py(4) bilayers. We have also measured the saturation magnetization $M_S = 6.4 \times 10^5$ A/m in test samples [17].

To verify the SHE origin of field-symmetric components of the FMR signals, we have studied several different types of control samples. In Fig. 2(c) we compare the FMR signals measured at 8 GHz for a Pt(15)/Py(15) and a Pt(6)/Py(4) sample. The signal for the Pt(6)/Py(4) sample contains a sizable field-symmetric component, with $S/A = 0.63$. Due to the increased thicknesses of the two layers, we expect from Eq. (3) that $S/A$ for the Pt (15)/Py (15) should be greatly reduced, approximately $\propto 1/td$ if the spin Hall currents in the two samples are similar. $S/A$ for the Pt(15)/Py(15) is very small, $S/A = 0.08 \pm 0.05$, near the noise floor for the fits of the symmetric component (the uncertainty reflects the standard deviation over five samples measured). The difference between the change in the $S/A$ ratio expected from Eq. (3) (a factor of 11.2, taking into account a small change in $4\pi M_{eff}$) and the measured reduction by a factor 8.0 may be associated with a change in the magnitude of the spin Hall current generated by the different thicknesses of the Pt films when this thickness is comparable to the spin diffusion length (see below).

We also studied control samples with the layers Cu(6)/Py(4) and 4 nm of Py alone, with results as shown in Fig. 2(d). The Cu/Py bilayer sample gives a purely antisymmetric signal, indicating that only the Oersted-field contribution is present, as expected because of the very small SHE in Cu in comparison to that in Pt. For the Py(4) sample, we would expect no resonance signal at all, since there is no SHE and as noted above if the current density in the Py is uniform there should also be no net effect of the Oersted field on the Py dynamics. However, we do observe a very small, purely antisymmetric signal in the 4 nm Py sample. We suspect that this may arise from an Oersted field due to non-uniform current flow at the ends of the Py due to the electrode contacts. The lack of field-symmetric components in the resonance curves for the control samples provides strong support that the symmetric component we observe in Pt(6)/Py(4) does indeed arise from the SHE in the Pt.

With $4\pi M_{eff}$ and $M_S$ determined, we can use Eq. (3) and the measured values of $S/A$ to calculate $J_{S,RF}/J_{C,RF}$. The results are shown in Fig. 2(e) for the resonance curves spanning 5-10 GHz shown in Fig. 2(a). We find $J_{S,RF}/J_{C,RF} = 0.056 \pm 0.005$ for Pt(6)/Py(4). We measured more than ten Pt(6)/Py(4) samples with different lateral dimensions and the total variation of $J_{S,RF}/J_{C,RF}$ was < 15%. The dominant experimental uncertainty [and the small variation with $H_{ext}$ visible in Fig. 2(e)] may be associated with Oersted fields from non-uniform currents at the sample ends, as noted above for the single-layer Py sample. Note that according to Eq. (2) $S/A$ should not depend upon the angle of the applied DC field, as confirmed by the results shown in Fig. 2(f).

As an independent check we also employed an alternative method for determining the spin current density absorbed by the Py layer, by measuring the FMR linewidth $\Delta$ as a function of DC current, similar to the technique introduced in Ref. 8. According to the theory of ST, a DC spin current $I_{S,DC}$ will increase or decrease the effective magnetic damping and hence $\Delta$, depending upon its relative orientation with respect to the magnetic moment: [18]

$$\Delta = \frac{2\pi f}{\gamma}\left(\alpha + \frac{\sin\theta}{(H_{ext}+2\pi M_{eff})\mu_0 M_S t}\frac{\hbar J_S}{2e}\right) \quad (4)$$

Our results obtained with a Pt(6)/Py(4) sample ~1 μm wide are shown in Fig. 3. The measured damping coefficient at zero current ($\alpha \approx 0.028$) is significantly higher than that measured in a spin valve nanopillar sample having a 4 nm Py free layer ($\alpha \approx 0.01$) [19]. This can be explained by the spin pumping effect previously observed in the Py/Pt system [20, 21]. For a negative applied field ($H_{ext}$ applied -135° from the current direction in the microstrip), the linewidth is broadened when $I_{DC}$ ramps from -0.7 mA to 0.7 mA; while for a positive field ($H_{ext}$ applied 45° from the current direction), the trend is the opposite. By fitting the data shown in Fig. 3, and calculating the charge current density in the Pt using the measured resistivities $\rho_{Pt} = 20 \mu\Omega cm$ and $\rho_{Py} = 45 \mu\Omega cm$, we find $\Delta\alpha/J_c = (0.9 \pm 0.012) \times 10^{-10}$ (A/cm$^2$)$^{-1}$. With Eq. (4), this yields $J_s/J_c = 0.048 \pm 0.007$ for Pt(6)/Py(4), which agrees well with the value $0.056 \pm 0.005$ determined from the FMR lineshape.

Our experiments yield values for $J_S/J_C$, the ratio of the spin current density (in units of charge) absorbed by the Py to the charge current density in the Pt film. For many applications, this is the figure of merit of direct interest. However, for comparing to other experiments, it is also of interest to determine the spin Hall angle $\theta_{SH}$, the ratio of the spin current density inside bulk Pt to the charge current density. For a perfectly transparent Pt/Py interface and for a Pt layer much thicker than the spin diffusion length $\lambda_{sf}$, the quantities $J_S/J_C$ and $\theta_{SH}$ should be equal. However, because our Pt/Py interface is likely not perfectly transparent, and because our Pt layers likely do not have thicknesses $\gg \lambda_{sf}$, our results may underestimate the transverse spin current density appropriate to bulk Pt. Therefore, our measurements imply a lower bound, $\theta_{SH} > 0.056 \pm 0.005$ for our Pt material. In the limit of a transparent Pt/Py interface, for which there should be no spin accumulation transverse to the Py moment at the interface, we calculate using drift-diffusion theory [22] that the spin Hall current density in a Pt film of thickness $d$ should be reduced from the bulk value by $J_S(d)/J_S(\infty) = 1 - \text{sech}(d/\lambda_{sf})$. Using this expression, our best estimate, based on comparison between the Pt(15)/Py(15) and Pt(6)/Py(4) samples is that $\lambda_{sf} \approx 3nm$, and we can set an upper bound of $\lambda_{sf} < 6nm$, lower than the low temperature value measured previously [23]. This gives a best estimate of $\theta_{SH} = 0.076$, and bounds $0.056 \pm 0.005 < \theta_{SH} < 0.16$, again in the limit of a transparent Pt/Py interface.

We mentioned above that previous measurements of $\theta_{SH}$ in Pt have differed by over an order of magnitude. Kimura *et al.* [9], using a Pt/Cu/Py lateral nonlocal geometry reported $\theta_{SH} = 0.0037$. However, their 4-nm-thick Pt wires are in contact to 80-nm-thick Cu wires. We believe that the Cu likely shunted the charge current flowing in the Pt, resulting in a large underestimation of $\theta_{SH}$. Ando *et al.* [8] by measuring magnetic damping in Pt/Py versus current, reported $J_S/J_C = 0.03$ and estimated $\theta_{SH} = 0.08$. We have shown that a technique closely related to the method of Ref. 8 gives results that agree with our FMR method, although we differ with Ref. 8 regarding the form of our Eq. (4) and the drift-diffusion analysis. Mosendz *et al.* [10], using a technique based on spin pumping together with the inverse SHE, reported $\theta_{SH} = 0.0067$, later refined to $\theta_{SH} = 0.013$ [24]. This result relied on an assumption that $\lambda_{sf} = 10$ nm for Pt. Their value for $\theta_{SH}$ would be 3 times larger, and in much better accord with our value, using our estimate that $\lambda_{sf} = 3$ nm.

In summary, we demonstrate that spin current generated by the SHE in a Pt film can be used to excite spin-torque FMR in an adjacent metallic ferromagnet (Py) thin film. This technique allows a straightforward determination of the efficiency of spin current generation, $J_s/J_c$ (the spin current density absorbed by the Py divided by the charge current density in the Pt), that is self-calibrated, in that the torque due to the spin current can be measured relative to the torque from the Oersted field generated by the same charge current density in the Pt layer. We find $J_s/J_c = 0.056 \pm 0.005$ for Pt(6)/Py(4), implying $\theta_{SH} > 0.056$ for bulk Pt. This simple technique is an excellent solution for the quantitative measurement of the SHE efficiency in any metallic film that can be produced as part of a ferromagnetic/non-magnetic metal bilayer. The relatively large efficiency of spin current generation that we observe for Py/Pt is promising for applications which might utilize the SHE to manipulate ferromagnet dynamics.

This research was supported in part by the Army Research Office, and by the NSF-NSEC program through the Cornell Center for Nanoscale Systems. This work was performed in part at the Cornell NanoScale Facility, which is supported by the NSF through the National Nanofabrication Infrastructure Network and benefitted from the use of the facilities of the Cornell Center for Materials Research, supported by the NSF-MRSEC program.

**FIGURE CAPTIONS**

Fig. 1 (color online): (a) Schematic of a Pt/Py bilayer thin film illustrating the spin transfer torque $\tau_{STT}$, the torque $\tau_H$ induced by the Oersted field $H_{RF}$, and the direction of the damping torque $\tau_\alpha$. $\theta$ denotes the angle between the magnetization $M$ and the microstrip. $H_{ext}$ is the applied external field. (b) Left side view of the Pt/Py system, with the solid line showing the Oersted field generated by the current flowing just in the Py layer, which should produce no net effect on the Py AMR. (c) Schematic circuit for the ST-FMR measurement.

Fig. 2 (color online): (a) Spectra of ST-FMR on a Pt(6)/Py(4) sample measured under frequencies of 5-10 GHz. The sample dimension is 20 μm wide × 110 μm long. Inset: ST-FMR spectrum of 8 GHz for both positive and negative $H_{ext}$. (b) Resonance frequency $f$ as a function of the resonant field $H_0$. The solid curve represents a fit to the Kittel formula. (c) FMR spectra measured for two Pt/Py bilayer samples, with fits to Eq. (2). The data were taken at 8 GHz. (d) FMR spectra ($f$ = 8 GHz) on the Pt(6)/Py(4) sample (blue triangles) as well as control samples consisting of Cu(6)/Py(4) (red circles) and Py(4) (black squares). (e) $J_{S,RF} / J_{C,RF}$ values determined from the FMR analysis [Eq. (3)] at different $f$. (f) FMR signals measured for different external field angles $\theta$ ($f$ = 8 GHz). The mixing voltages $V_{mix}$ are normalized and offset to enable comparison of the lineshapes.

Fig.3 (color online): The change of the FMR linewidth (left y axis) and Gilbert damping coefficient (right y axis) as a function of $I_{DC}$ for two orientations of the Py magnetization relative to the current direction. The data are taken at $f$ = 8 GHz.

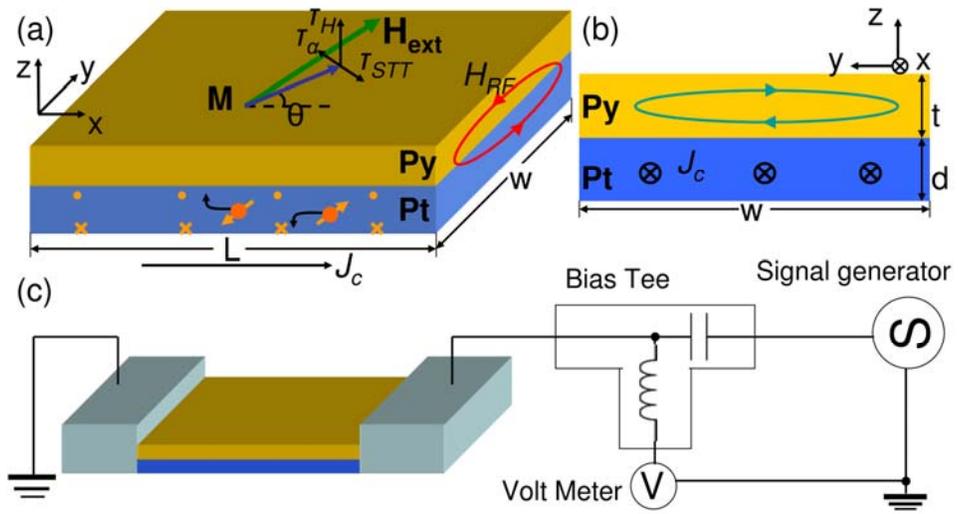

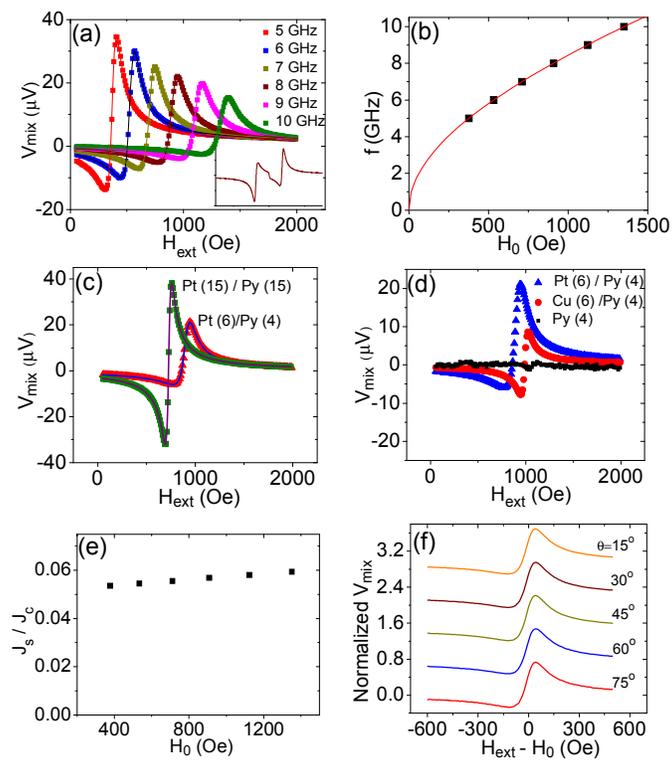

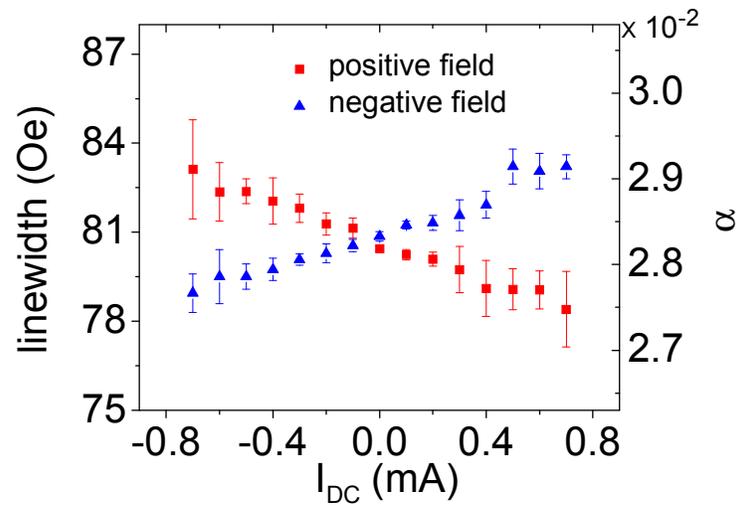